\documentclass[
  twocolumn,
  prl,
  showpacs,
  amsmath,
  amssymb,
  superscriptaddress,
  floatfix
]{revtex4}

\usepackage{bm}
\usepackage{dsfont}
\usepackage{graphicx}

\newcommand{\s}{\sum\limits}

\newcommand{\pa}{\partial}

\newcommand{\be}{\begin{equation}}
\newcommand{\e}{\end{equation}}
\newcommand{\beml}{\begin{subequations}}
\newcommand{\eml}{\end{subequations}}
\newcommand{\beq}{\begin{eqnarray}}
\newcommand{\eq}{\end{eqnarray}}
\newcommand{\ba}{\begin{array}}
\newcommand{\ea}{\end{array}}
\newcommand{\bpm}{\begin{pmatrix}}
\newcommand{\epm}{\end{pmatrix}}
\newcommand{\lt}{\left}
\newcommand{\rt}{\right}

\newcommand{\la}{\langle}
\newcommand{\ra}{\rangle}
\newcommand{\ep}{\varepsilon}

\newcommand{\bb}{\mathbf}
\newcommand{\bs}{\boldsymbol}

\DeclareMathOperator{\tr}{Tr}

\DeclareMathOperator{\re}{Re}

\begin{document}

\title{Color-dependent conductance of graphene with adatoms}

\author{J.~Schelter}
\affiliation{
 Institut f\"ur Theoretische Physik und Astrophysik, University of W\"urzburg, 97074
 W\"urzburg, Germany
}

\author{P.~M.~Ostrovsky}
\affiliation{
 Institut f\"ur Nanotechnologie, Karlsruhe Institute of Technology,
 76021 Karlsruhe, Germany
}
\affiliation{
 L.~D.~Landau Institute for Theoretical Physics RAS,
 119334 Moscow, Russia
}

\author{I.~V.~Gornyi}
\affiliation{
 Institut f\"ur Nanotechnologie, Karlsruhe Institute of Technology,
 76021 Karlsruhe, Germany
}
\affiliation{
 A.F.~Ioffe Physico-Technical Institute,
 194021 St.~Petersburg, Russia.
}

\author{B.~Trauzettel}
\affiliation{
 Institut f\"ur Theoretische Physik und Astrophysik, University of W\"urzburg, 97074
 W\"urzburg, Germany
}

\author{M.~Titov}
\affiliation{
 School of Engineering \& Physical Sciences, Heriot-Watt University,
 Edinburgh EH14 4AS, UK
}
\affiliation{
 Institut f\"ur Nanotechnologie, Karlsruhe Institute of Technology,
 76021 Karlsruhe, Germany
}

\begin{abstract}
We study ballistic transport properties of graphene with a low concentration of vacancies or adatoms. The conductance of graphene doped to the Dirac point is found to depend on the relative distribution of impurities among different sites of the honeycomb lattice labeled in general by six colors. The conductivity is shown to be sensitive to the crystal orientation if adatom sites have a preferred color. Our theory is confirmed by numerical simulations using recursive Green's functions with no adjustable parameters.
\end{abstract}


\maketitle

Controllable deposition of adatoms known as chemical functionalization is an efficient way to alter graphene's electronic properties. Adatoms can change the orbital state of the functionalized carbon atoms from $sp^2$ to $sp^3$ configuration removing electrons from the conduction band and transforming graphene into a semiconductor \cite{Liu08,Elias09}. For instance, fully hydrogenated graphene (graphane) is predicted to have a wide band gap of $3-14$ eV \cite{Lebegue10} in sharp contrast to clean graphene which has a gapless excitation spectrum \cite{Novoselov04}.

A small concentration of strongly bound adatoms or molecules, such as hydrogen or CH$_3$, is naturally present even in the cleanest graphene samples produced by exfoliation. The effect of vacancies, which can be thought of as infinitely strong on-site potential impurities, is essentially analogous to that of adatoms. Such impurities support electronic states at the Dirac point, yielding resonant scattering of Dirac quasiparticles
\cite{Ostrovsky06,Stauber07,Pereira08,Robinson08,Elias09Ni10Katoch10,Titov10,Ostrovsky10,Wehling10}.

From a theoretical point of view, the strong on-site impurities or vacancies preserve both time-reversal and chiral symmetries of the graphene Hamiltonian. The interplay of the symmetries prevents quantum localization at the Dirac point \cite{Ostrovsky06} and gives rise to a quantum critical regime of charge transport at sufficiently high impurity concentration. The conductivity of graphene in this regime is predicted to take on a constant value, which depends on the distribution of adatoms among different sublattices of the graphene crystal \cite{Ostrovsky10}. In this Letter, we discover an even more subtle sensitivity of transport quantities, not only to sublattice but also to the type of impurity sites therein (described by the site's ``color''), that can be clearly seen even in the presence of two scatterers.

%%%%%%%%%%%%%%%%%%%%%%%%%%%%
%%%% fig:color
%%%%%%%%%%%%%%%%%%%%%%%%%%%%
\begin{figure}
\centerline{\includegraphics[width=0.8\columnwidth]{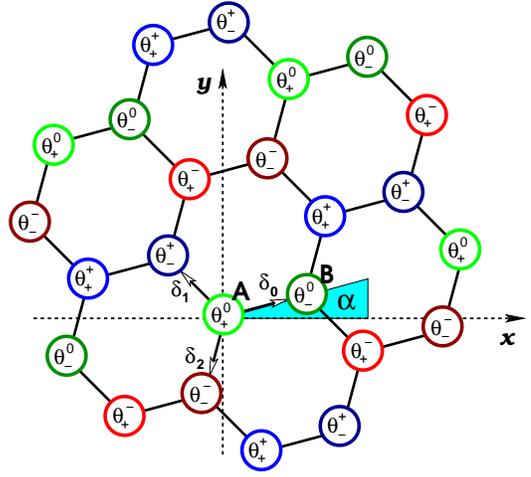}}
\caption{(Color online) Color scheme for vacancies or resonant adatoms. The impurity site is characterized by the phase $\theta_{\pm}^{c}=\pm\alpha+2\pi c/3$, where $\pm$ refers to the sublattice (A or B) and $c =-1,0,1$ denotes the colors (red, green, blue). The color scheme depends on the transport direction $x$.  The phases at $A$ and $B$ sites connected by the vector $\bs{\delta}_0$ are equal for $\alpha=0$, while they differ most for $\alpha=\pi/6$.
}
\label{fig:color}
\end{figure}
%%%%%%%%%%%%%%%%%%%%%%%%%%%%%

The theory of quantum transport in disordered graphene is highly non-trivial in the vicinity of the Dirac point due to the breakdown of the quasiclassical approximation. An essentially exact approach, which is referred to below as an unfolded scattering formalism, has been proposed to tackle the problem in Ref.\ \cite{Ostrovsky10}. In this paper, we perform for the first time a detailed comparison of this formalism with numerical simulations based on the well-established recursive Green's function technique. In particular, we calculate both numerically and analytically the conductance of graphene at the Dirac point in the presence of two vacancies. This quantity is shown to have a remarkable sensitivity to the ``color'' of the vacated sites, which is determined by the phase of the Bloch wave-function and by the orientation of the crystal with respect to the transport direction. We further use the analytic results in order to construct the first terms of the color-dependent virial expansion of the conductivity with respect to vacancy concentration.

An essential ingredient of the analytical approach developed in Refs.\ \cite{Titov10, Ostrovsky10} is a single-impurity $T$-matrix, which is known for a variety of impurity types in graphene \cite{HentschelNovikovBasko} and carbon nanotubes \cite{Mccann04}. Starting from the standard tight-binding model on the honeycomb lattice, one can relate the onsite potential $V_i$ to the corresponding $T$-matrix by the geometric series $T_i = V_i + V_i g_{ii} V_i + V_i g_{ii} V_i g_{ii} V_i + \dots$, where $g_{ij}$ is the free particle Green's function taken between the lattice sites $i$ and $j$. The key simplification occurs at the Dirac point ($\ep=0$) since the Green's function element $g_{ii} \propto \ep \ln \ep $ vanishes in the limit $\ep\to 0$.  The vanishing amplitude of return yields at $\ep=0$ a simple form of the $T$-matrix operator: $T_i=V_i.$ Expressing this $T$-matrix in the valley-sublattice space of the Dirac Hamiltonian reveals a color scheme for the graphene lattice, which we shall describe below in detail.

To be more specific we write down the tight-binding model of ideal graphene as
\be
-t \sum_s \Psi(\bb{r} + \zeta_{\bb{r}} {\bs{\delta}}_s)
=\ep \Psi(\bb{r}),
\e
where the sign factor $\zeta_{\bb{r}}=\pm$ specifies the sublattice of the $\bb{r}$ site ($A$ or $B$), $t \approx 2.7$eV is the hopping integral, and the index $s$ takes on three values: $0$, $1$, and $2$. We regard $x$ as transport direction and orient the graphene lattice at an arbitrary angle $\alpha$ as shown in Fig.~\ref{fig:color}. The sites of the honeycomb lattice are then connected by three vectors
\be
\bs{\delta}_s = a \bpm \cos(\alpha+2\pi s/3) \\ \sin(\alpha+2\pi s/3) \epm,\qquad s=0,1,2,
\e
where $a$ is the distance between the neighboring carbon atoms.
The two non-equivalent Dirac-point vectors in the reciprocal space, $\bb{K}=(4\pi/3a\sqrt{3}) (\sin \alpha, -\cos\alpha)$ and $\bb{K}'=-\bb{K}$, define the two valleys. The effective mass model of graphene is introduced by the ansatz
\be
\label{Psi_lat}
\Psi(\bb{r}) = \sqrt{A} \lt[ e^{i \bb{K r} } \phi_{\zeta_{\bb{r}}} + e^{-i\bb{K r}}\phi'_{\zeta_{\bb{r}}} \rt],
\e
where two pairs of envelope functions $\phi_\pm$ and $\phi'_\pm$, that correspond to the valleys $\bb{K}$ and $\bb{K}'$, are smooth on the scale $a$ and $A = (3\sqrt{3}/2) a^2$ is the area of the elementary cell. Arranging these four slowly varying amplitudes in a single vector
\be
\label{Phi}
|\Phi\ra=\Big( e^{-\tfrac{i\alpha}{2}} \phi_+ , -i e^{\tfrac{i\alpha}{2}} \phi_- , -i e^{-\tfrac{i\alpha}{2}} \phi'_- , e^{\tfrac{i\alpha}{2}} \phi'_+  \Big)^T,
\e
one arrives at the valley-symmetric Dirac Hamiltonian, $H_0= - i \hbar v \boldsymbol{\sigma \nabla}$, where  $\hbar v = 3 t a /2$ and $\boldsymbol{\sigma}=(\sigma_x,\sigma_y)$ is the vector of Pauli matrices in the sublattice space.

The wave function on the honeycomb lattice (\ref{Psi_lat}) is expressed as $\Psi(\bb{r}) = \la u(\bb{r}) | \Phi \ra$ where
\beq
\label{u}
\la u(\bb{r}) | = \sqrt{A} \begin{cases}
\Big( e^{\tfrac{i\alpha}{2} + i \bb{K r} } , 0 , 0 , e^{-\tfrac{i\alpha}{2} -i\bb{K r}} \Big), & \zeta_{\bb{r}} = +,\\
\Big( 0 , i e^{-\tfrac{i\alpha}{2}+i \bb{K r}} , i e^{\tfrac{i\alpha}{2} - i\bb{K r}} , 0 \Big), & \zeta_{\bb{r}} = -.
\end{cases}
\eq
The representation of the $T$-matrix of the on-site potential impurity located at $\bb{r}$ is given by \cite{Mccann04}
\be
\label{Tmat}
T =  \lt| u(\bb{r}) \ra V \la u(\bb{r}) \rt|.
\e

The phase factors in Eq.\ (\ref{u}) are responsible for a site classification, which depends on the orientation angle $\alpha$. Using Eqs.\ (\ref{u}, \ref{Tmat}) we represent the impurity $T$-matrix as
\be
\label{T}
T_\zeta = \ell\lt(1+\zeta\sigma_z\tau_z+\sigma_{-\zeta}\tau_-e^{i\theta_\zeta}+\sigma_\zeta\tau_+e^{-i\theta_\zeta}\rt),
\e
where $\ell=A V/2$ is the scattering length, $\sigma_\pm=(\sigma_x\pm i\sigma_y)/\sqrt{2}$, $\tau_\pm=(\tau_x\pm i\tau_y)/\sqrt{2}$, $\sigma_{x,y,z}$ and $\tau_{x,y,z}$ are the Pauli matrices in the sublattice and valley space, respectively. The phases are $\theta_\pm = \pm \alpha + 2 \bb{K r}$. Assuming an A site at $\bb{r} = 0$, we encounter three possible values of the factor $e^{2i\bb{K r}} = e^{-i\bb{K r}} = e^{2\pi i c/3}$ with $c=0,\pm 1$. Together with the sublattice index $\zeta$ this yields six possible $T$-matrices for on-site impurities. In order to visualize this classification, let us introduce a color scheme by assigning six colors to lattice sites (three in each sublattice) corresponding to the phase factors $e^{i \theta_\pm^c} = e^{\pm i \alpha + 2\pi i c/3}$. These colors define a superlattice (shown in Fig.\ \ref{fig:color}) with a period of three elementary unit cells and with six atoms per supercell. For $\alpha=0$, only three distinct phases remain, yielding the same three colors in both sublattices.

\begin{figure}
\centerline{\includegraphics[width=0.9\columnwidth]{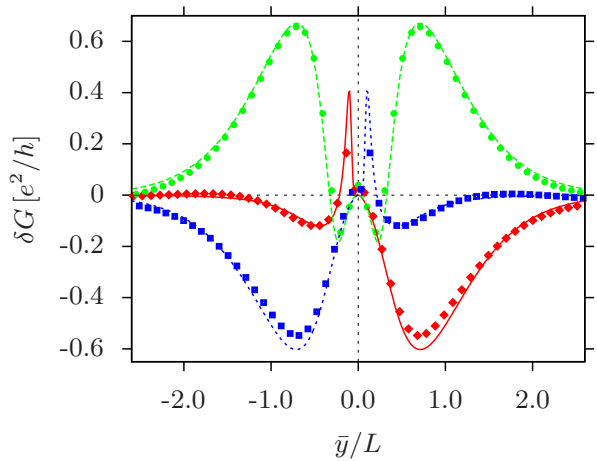}}
\caption{(Color online) Conductance variation for an ``armchair'' ($\alpha=0$) sample with two vacancies (A and B). Changing the distance $\bar{y}$, the conductance jumps on the atomic scale between three different smooth curves corresponding to $\bar\theta = 0$ (green disks), $\bar\theta = 2\pi/3$ (blue squares), and $\bar\theta = -2\pi/3$ (red diamonds). The numerical data agrees well with Eq.\ (\ref{2result}), as shown by the corresponding curves.
}
\label{fig:armchair}
\end{figure}

\begin{figure}
\centerline{\includegraphics[width=0.8\columnwidth]{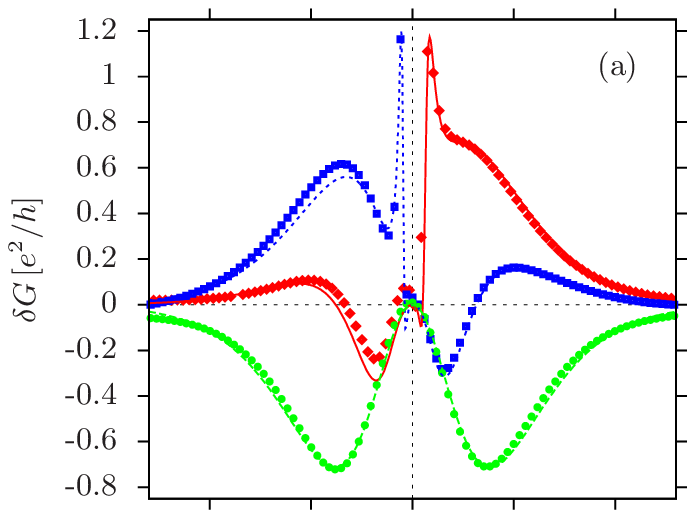}}
\centerline{\includegraphics[width=0.8\columnwidth]{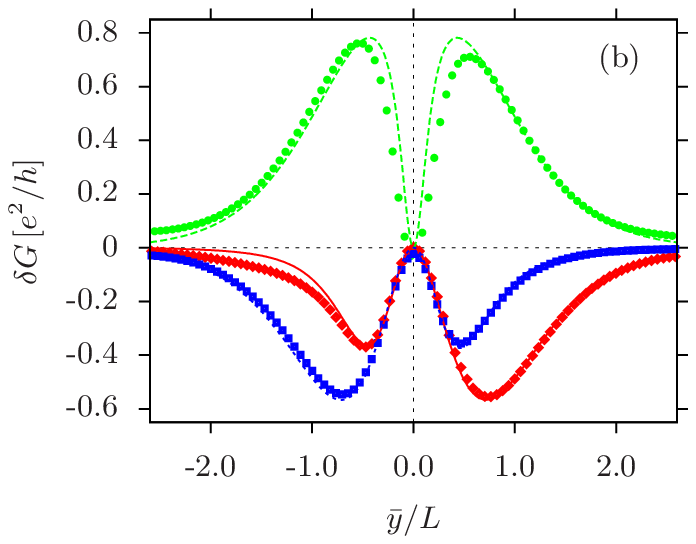}}
\caption{(Color online) Conductance variation for a ``zigzag'' ($\alpha=\pi/6$) sample with two vacancies. Top panel: A and B vacancies.
On changing $\bar{y}$ the conductance jumps on the atomic scale between different smooth curves corresponding to $\bar{\theta}=\pi$ (green disks), $\pi/3$ (blue squares), and $-\pi/3$ (red diamonds). Bottom panel: Both vacancies in A sublattice. The phase difference $\bar\theta$ is either  $0$ (green disks), $2\pi/3$ (blue squares), or $-2\pi/3$ (red diamonds). The numerical data agrees well with the result Eq.\ (\ref{2result}), as shown by the corresponding curves.}
\label{fig:zigzag}
\end{figure}

In order to demonstrate the sensitivity of transport quantities to the colors of impurity sites, we compute the conductance of a rectangular graphene sample  with two vacancies at positions $\bb{r}_1$ and $\bb{r}_2$. The conductance is calculated numerically using a modification \cite{Schelter10} of the recursive Green's functions algorithm \cite{RGFreferences} that accounts for peculiarities stemming from the graphene lattice structure,
as well as analytically from the unfolded scattering formalism \cite{Ostrovsky10}. The numerical and analytical results are compared in Figs.~\ref{fig:armchair}, \ref{fig:zigzag}.

%The general theory of Ref.\ \cite{Ostrovsky10} can be applied to calculate the conductance of graphene with $N$ localized impurities (with non-overlapping potentials), which are specified by their $T$-matrices and coordinates. The key advantage of the theory is the reduction of a microscopic calculation of transport and spectral quantities to a finite-size matrix algebra in the unfolded (impurity) space.

We consider a rectangular graphene sample of the dimensions $L \times W$ with metallic leads attached at $x=0$ and $x=L$. In order to model the pseudodiffusive transport regime, we assume $W \gg L$. In our analytic approach, the chemical potential in the leads is fixed infinitely far from the Dirac point, while the chemical potential inside the sample is set to zero. For two on-site impurities with scattering length $\ell$, the correction to the conductance as compared to the clean case is conveniently expressed as
\be
\label{conductance}
\delta G=\frac{4e^2}{h} \lt\{ \frac{\pa^2}{\pa \phi^2} \ln \det \lt[ \mathds{1} -\ell\bpm
\mathcal{G}_{11} & \mathcal{G}_{12} \\
\mathcal{G}_{21} & \mathcal{G}_{22}
\epm\rt]\rt\}_{\phi=0},
\e
where $\mathcal{G}_{ij} = \la u(\bb{r}_i) | \hat{\mathcal{G}} | u(\bb{r}_j) \ra $ are the matrix elements of the Green's function operator defined by
\begin{equation}
 \label{keldysh}
 \begin{pmatrix}
   \mu - H_0 + i0 & -\sigma_x \eta \delta(x) \\
   -\sigma_x \eta \delta(x-L) & \mu - H_0 - i0
 \end{pmatrix}\hat{\mathcal{G}}=\hat{\mathds{1}}.
\end{equation}
Here $\eta = i \sinh(\phi/2)$ is the counting field applied at the metal-graphene interfaces $x=0$ and $x=L$. Below we study the case of two vacancies modeled by an infinite on-site potential. This corresponds to the limit $\ell \to \infty$ and allows us to simplify Eq.\ (\ref{conductance}).

The exact analytical solution to Eq.~\eqref{keldysh} with the specified choice of the chemical potential was found in Ref.\ \cite{Titov10}. Using this solution together with the expressions (\ref{u}) in Eq.~\eqref{conductance}  we express the conductance of a rectangular graphene sample with two vacancies as $G=G_0+\delta G$, where $G_0=4e^2 W/\pi h L$ is the conductance of the clean sample. The vacancies located at positions $\bb{r}_1$ and $\bb{r_2}$ inside the sample give rise to the correction
\be
\label{2result}
\delta G=\frac{4e^2}{h}\, \frac{\bar{y}^2}{L^2} \,
\frac{\rho_{11}\rho_{22} \re (\rho_{12}^2)-|\rho_{12}|^2 (\re \rho_{12})^2}
{\big[(\re \rho_{12})^2-\rho_{11}\rho_{22}\big]^2},
\e
where $\bar{y}=y_1-y_2$. Here we introduced the notation
\be
\rho_{ij} = e^{i (\theta_i-\theta_j )/2}\csc\big[\pi(\zeta_ix_i+\zeta_jx_j+i y_i-i y_j)/2L\big],
\e
where $\theta_i=\alpha \zeta_i - \bb{K}\bb{r}_i$ is the $T$-matrix phase, which can take on one
of six values $\theta_{\pm}^c$. The phase difference $\bar{\theta}=\theta_1-\theta_2$ encodes
the dependence of conductance on the colors of vacancies and on the orientation angle $\alpha$.

For numerical simulations we have chosen the sample length $L\approx 50\, a$, the width $W \approx 600\, a,$ and consider two crystal orientations with $\alpha=0$ (armchair) and $\alpha=\pi/6$ (zigzag). The metallic leads at $x < 0$ and $x > L$  are defined by the large chemical potential, $\mu_\infty  = 0.3\, t \approx  10\, \hbar v/L$ (measured with respect to the Dirac point).  Inside the sample ($0 < x < L$) the chemical potential is tuned to the close vicinity of the Dirac point, $\mu_0 = 0.001\, t \approx  0.033\, \hbar v/L$. In the absence of vacancies, the relative deviation of the conductance from the value $G_0= 4e^2 W/ \pi h L$ is found to be less than $1\%$. While a single vacancy has only a negligible effect on the numerical results, a pair of vacancies leads to a finite correction, $\delta G$, which we calculate numerically for different positions of one vacancy while keeping the second vacancy fixed in the center of the sample.

One set of data is shown in Fig.\ \ref{fig:armchair} for $\alpha=0$ and for vacancies of different type: $A$-vacancy at $\bb{r}_1=(L/2,0)$ and $B$-vacancy at $\bb{r}_2=(x_2,y_2)$ where $x_2 \approx 2L/3$. The dependence of $\delta G$ on all possible values of $\bar{y}=-y_2$ on the lattice is shown with the data points. The conductance switches between three different smooth curves which correspond to a certain difference, $\bar{\theta}=\theta_1-\theta_2$, of the $T$-matrix phase at the two vacated sites, $\bar{\theta}=0$ and $\bar{\theta}=\pm 2\pi/3$, respectively. Similar data is plotted in Fig.\ \ref{fig:zigzag}a for the zigzag orientation of the crystal and for different positions of the $B$-vacancy, $x_2 \approx 0.45L$. Here three other phase differences, $\bar{\theta}=\pi$ and $\bar{\theta}=\pm\pi/3$, appear. In Fig.~\ref{fig:zigzag}b both vacancies are chosen to belong to the $A$ sublattice ($x_2 \approx L/3$, zigzag orientation). In this case we find again $\bar{\theta}=0$ and $\bar{\theta}=\pm 2\pi/3$ in agreement with the classification of site colors presented in Fig.\ \ref{fig:color}.

%%%%%%%%%%%%%%%%%%%%%%%%%%%%
%%%% fig:color
%%%%%%%%%%%%%%%%%%%%%%%%%%%%
%%%%%%%%%%%%%%%%%%%%%%%%%%%%%

In Figs.\ \ref{fig:armchair}, \ref{fig:zigzag} we demonstrate a remarkable agreement between the numerical data obtained by the recursive Green's function technique and the analytical result (\ref{2result}). The small deviations between analytical and numerical results are mainly due to the finite ratio $W/L$ and the chosen finite chemical potential in the leads; a small detuning from the Dirac point within the sample, used in the simulations to avoid numerical instabilities, may also play a role. A good agreement with the theory is similarly obtained for other positions of the vacancies.

Transport in the ``short and wide'' ($W \gg L$) sample can be described by the mean conductivity $\sigma=L G/W$. For a low concentration of vacancies a virial expansion for $\sigma$ can be established. The lowest (second) order term of this series is readily found from Eq.\ \eqref{2result}:
\begin{multline}
\!\!\!\sigma =\frac{4e^2}{\pi h} \bigg\{ 1 + L^4\! \s_{c_1, c_2} \Big[
\gamma_{\frac{2\pi}{3}(c_1-c_2)} (n_{Ac_1} n_{Ac_2} + n_{Bc_1} n_{Bc_2}) \\
+2 \bar\gamma_{2\alpha + \frac{2\pi}{3}(c_1-c_2)} n_{Ac_1} n_{Bc_2} \Big]
+\mathcal{O}(n^3L^6) \bigg\},
\label{sigma}
\end{multline}
where $n_{Ac}$ and $n_{Bc}$ with $c=0,\pm 1$ are the impurity concentrations of the corresponding sites of the graphene superlattice (Fig.\ \ref{fig:color}). Performing numerical averaging over impurity positions in Eq.\ \eqref{2result} we find the values $\gamma_0 \approx 0.2653$ and $\gamma_{\pm 2\pi/3} \approx -0.1197$, which are independent of the angle $\alpha$. Both $\gamma_\chi$ and $\bar{\gamma}_\chi$ are even $2\pi$-periodic functions of $\chi$. The mean value of $\bar{\gamma}_\chi$ over the period is zero. In a sample with ``armchair'' orientation ($\alpha = 0$) relevant parameters are $\bar\gamma_0 \approx 0.1700$ and $\bar\gamma_{\pm 2\pi/3} \approx -0.0850$. In a ``zigzag'' sample ($\alpha = \pi/6$), we have $\bar\gamma_{\pm \pi/3} \approx 0.0843$ and $\bar\gamma_\pi \approx -0.1686$. In each case, the sum of the three values is vanishingly small $\sim 10^{-5}$. If impurities are distributed uniformly among the lattice sites, the conductivity is given by
\be
\sigma=\frac{4e^2}{\pi h} \,\lt(1+ \kappa \, n^2L^4 + \mathcal{O}(n^3L^6)\rt),
\e
where $n$ is the impurity concentration and $\kappa \approx 0.0043$ exhibits tiny oscillations $\sim 10^{-5}$ as a function of $\alpha$.

%%%%%%%%%%%%%%%%%%%%%%%%%%%%
%%%% fig:color
%%%%%%%%%%%%%%%%%%%%%%%%%%%%
%%%%%%%%%%%%%%%%%%%%%%%%%%%%%

However, the dependence of the conductivity on the crystal orientation is substantial if the distribution of adatoms among sites of different colors is not uniform. Such a non-uniform distribution can originate from adatom correlations. For instance, the electron-mediated interaction of on-site impurities \cite{Shytov09, Abanin10} is inferred from the free energy $F=-T\sum_{\omega_n}\tr \ln (1-\ell\la u|\hat{\mathcal{G}}| u \ra)$, where $\hat{\mathcal{G}}$ is the Green's function (\ref{keldysh}) at Matsubara energy $i\omega_n$ and $\phi=0$. This gives rise to interaction oscillations yielding a color scheme similar to that depicted in Fig.\ \ref{fig:color}.
It is worth noting that Refs. \cite{Shytov09} addressed the problem in an infinite system, whereas the effective interaction of adatoms with the leads
[encoded in Eq.~(\ref{keldysh})] may play an important role in a finite sample.

In conclusion, we theoretically proposed and numerically confirmed an extended classification of impurity sites in the graphene honeycomb lattice for the case of strongly-bound adatoms or vacancies. The classification is illustrated in Fig.~\ref{fig:color} by assigning colors to the lattice sites. The general analytical expression for the Dirac-point conductance of a graphene sample with two resonant on-site impurities is given as a function of impurity coordinates.  The Dirac-point conductivity of graphene with a small number of randomly distributed adatoms is shown to be sensitive to the relative concentration of impurities at the sites belonging to different sublattices and having different colors.

We are grateful to A. Mirlin and A. Shytov for stimulating discussions.
The support of the British Council and German Academic Exchange Service under ARC project 1381 is acknowledged.
The work was also supported by the DFG (in particular, by SPP ``Graphene''), by DFG -- Center for
Functional Nanostructures, by the DFG--RFBR cooperation grant, and by the EUROHORCS/ESF EURYI scheme (I.V.G.).

\end{document}